\documentclass[aps,twocolumn]{revtex4}
\usepackage{pgf,xcolor,tikz}

\begin{document}


\title{Thermal dissipation in two dimensional relativistic Fermi \\gases with a relaxation time model}


\author{A. R. M\'endez}
\email[]{amendez@correo.cua.uam.mx}
\author{A. L. Garc\'ia-Perciante}
\email[]{algarcia@correo.cua.uam.mx}
\author{G. Chac\'on-Acosta}
\email[]{gchacon@correo.cua.uam.mx}
\affiliation{Departamento de Matem\'aticas Aplicadas y Sistemas,\\ Universidad Aut\'onoma Metropolitana - Cuajimalpa, 
\\05348, Cuajimalpa, M\'exico.}

\date{\today}

\begin{abstract}\label{abs}
The thermal transport properties of a two dimensional Fermi gas are
explored, for the full range of temperatures and densities. The heat
flux is established by solving the Uehling-Uhlebeck equation using
a relaxation approximation given by Marle's collisional kernel and
considering the temperature and chemical potential gradients as independent
thermodynamic forces. It is shown that the corresponding transport
coefficients are proportional to each other, which leads to the possibility
of defining a generalized thermal force and a single transport coefficient.
The behavior of such conductivity with the temperature and chemical
potential is analyzed and a discussion on its dependence with the
relaxation parameter is also included. The relevance and applications
of the results are briefly addressed.  
\end{abstract}


\maketitle

\section{Introduction\label{intro}}
After the discovery of graphene \cite{Nat2005} the study of two-dimensional
statistical systems has acquired particular importance. Indeed, such
material is considered to be the most perfect two-dimensional electronic
system attainable, having exactly one thick atomic monolayer. All
the dynamics and transport processes are carried out confined to that
2D layer \cite{Peres}, so it can be studied as a two-dimensional
gas of relativistic fermions, where the Fermi energy imposes a limit
scale on the system \cite{nano}. Moreover, this phenomenon has been
successfully applied to the study of certain processes in relativistic
gases that can not be otherwise explored, in particular in an experimental
setting \cite{RBMG,SciRep2012}. 

Theoretical and experimental studies of non-relativistic fermion gases
in two or quasi-two dimensions, indicate that dimensionality plays
a role in the transport properties of the system. The corresponding
modifications have been explored by several authors \cite{lyakov,dellano,schafer,prl,valid,cheng,dyke}.
Certainly, by controlling the size of the systems it has been possible
to effectively manipulate the electronic properties of some materials
\cite{pnas}. 

The thermodynamic features of relativistic fermionic systems have
been also shown to be affected by a reduction in the number of dimensions
\cite{Blas,Cai}. For instance, different behaviors of the thermal
variables have been found at different temperature regimes depending
on the dimensionality \cite{Sevilla}. Indeed it can be seen by inspection
of the distribution function of a relativistic non-degenerate gas
in any dimension $d$, that for even values of $d$ the modified Bessel
functions that appear are of semi-integer index \cite{Chacon}, such
that it admits polynomial expansions \cite{Table}.

Moreover, for relativistic gases there is a threshold temperature
that depends on both dimensionality and statistics, from which relativistic
effects begin to be relevant in the description of the gas. This describes
a transition in the shape of the relativistic distribution: from unimodal
in the non-relativistic case, to bimodal in the ultra relativistic
limit \cite{SciRep2012}. This process has been characterized within
the theory of phase transitions and the corresponding critical exponents,
critical temperature and diverse thermodynamic properties have been
studied \cite{Morpho,Victoria}.

In spite of recent developments and vast motivation for the study
of non-equilibrium two-dimensional relativistic gases only a reduced
number of theoretical works on the matter can be currently found in
the literature. Indeed, relativistic kinetic theory for 2D systems
was first applied to specific cosmological models \cite{KD}, and
has also served to explore computational methods for heat flux \cite{Ghodrat}.
Hydrodynamic electron transport models have been recently studied
for graphene and other quasi-two dimensional materials (see for example
\cite{SciRep2013,Svintsov1,Svintsov2,Lucas}).

The transport coefficients of a 2D ultra-relativistic fluid were calculated
by Mendoza et. al. in \cite{UR2D} by introducing BGK-like relaxation
approximations for the collision operator and considering the Jüttner
distribution function for the equilibrium state. Whereas in \cite{HFBidi},
the constitutive equation for the heat flux and the corresponding
thermal conductivity were obtained for a non-degenerate gas by using
the complete collision term of the 2D relativistic Boltzmann equation
within the well-known Chapman-Enskog method. 

Also, and since in the hydrodynamic regime the electrons in graphene
can be thought of as a two-dimensional relativistic gas, a numerical
scheme has been constructed to study such electronic transport \cite{Oettinger}.
The condition for the existence of a Rayleigh-Benard instability was
studied in this framework and the thermal conductivity and the shear
viscosity were numerically established \cite{Rayleigh-Bernard}. A
lattice Boltzmann method has also beed recently developed for 2D relativistic
fluids with different statistics based on a fifth order expansion.
With this scheme the transport coefficients were numerically obtained
and corroborated against theoretical results and previous developments
\cite{Coelho2,Coelho}.

It is noteworthy that even in one and three dimensions, relativistic
dissipation and transport processes are still not fully understood.
Recent theoretical and numerical works in relativistic kinetic theory
strive to gain a deeper comprehension of such processes. These range
from comparisons between the Chapman-Enskog and Grad solution methods,
to the development and application of lattice Boltzmann methods and
other hydrodynamic solvers in the relativistic regime \cite{D,B,G,Gabbana,Ambrus}.

In this paper, the constitutive equation for the heat flux and the
corresponding coefficient of thermal conductivity for a two-dimensional
relativistic fermionic gas are established by solving the relativistic
Boltzmann equation for a degenerate Fermi gas considering a relaxation
approximation. In order to accomplish such task, the rest of the work
is structured as follows: In Section 2 we review the equilibrium solution
to the relativistic Ueling-Ulenbeck equation for fermions. The Fermi
energy is introduced by means of the statistical definition of the
number density and the corresponding relevant limits are briefly described.
In Section 3 the balance equations are addressed and the Chapman-Enskog
procedure is introduced, by means of which the local equilibrium equations
are obtained and the general expression for the heat flux is established
in terms of the first order out of equilibrium solution. The first
non-equilibrium correction to the distribution function is established
in Section 4 by replacing the collision operator of the relativistic
Ueling-Ulenbeck equation by a relaxation term, corresponding specifically
to the so-called Marle model. With such solution, the heat flux is
obtained in Section 5 by considering the temperature and chemical
potential gradients as independent thermodynamic forces. The discussion
section, Section 6, includes both a comparison with the 3D case and
relevant comments concerning the choice of relaxation time while Section
7 is devoted to the concluding remarks. 

\section{The Uehling-Uhlebeck equation and equilibrium solution for Fermions}

The kinetic description of a quantum system is given by the relativistic
Uehling-Uhlenbeck equation \cite{UU,KremerLibro} 
\begin{widetext}
\begin{equation}
p^{\alpha}\frac{\partial f}{\partial x^{\alpha}}=\int\left[\tilde{f}'f'\left(1-\frac{h^{2}}{g_{s}}f\right)\left(1-\frac{h^{2}}{g_{s}}\tilde{f}\right)-\tilde{f}f\left(1-\frac{h^{2}}{g_{s}}f'\right)\left(1-\frac{h^{2}}{g_{s}}\tilde{f}'\right)\right]F\sigma d\Omega\frac{d^{2}\tilde{p}}{\tilde{p}_{0}},\label{eq:Uhlebeck}
\end{equation}
\end{widetext}
which specifies the time evolution of the distribution function $f$,
i.e. the energy level occupation number. The particle coordinates
are here denoted by $\left(x^{\alpha}\right)=\left(ct,\,x^{1},\,x^{2}\right)$
and the corresponding momentum tensor by $\left(p^{\alpha}\right)=\left(p^{0},\,p^{1},\,p^{2}\right)$.
Space-time is specified by the metric $ds^{2}=\eta_{\alpha\beta}dx^{\alpha}dx^{\beta}$,
where $\alpha,\beta=0,\,1,\,2$ and $\eta_{\alpha\beta}=\text{diag}(1,-1,-1)$.
In Eq. (\ref{eq:Uhlebeck}), $F$ is the so-called invariant flux,
$\sigma$ the differential cross section and $\Omega$ is the solid
angle. Also, $h$ is Planck's constant and $g_{s}$ is the spin degeneracy
factor. Primed and unprimed quantities indicate values for the particles
before and after a collision respectively. 

The local equilibrium solution to the relativistic Uehling-Uhlenbeck
equation for particles that obey Pauli's exclusion principle (\ref{eq:Uhlebeck})
is the relativistic Fermi-Dirac distribution function, given by 
\begin{equation}
f^{(0)}=\frac{g_{s}}{h^{2}}\left(\Sigma+1\right)^{-1},\label{eq:f0Fermi}
\end{equation}
where we have introduced $\Sigma=\exp\left(-\mu_{e}/kT+u^{\alpha}p_{\alpha}/kT\right)$
in order to simplify the notation. Also here $k$ is the Boltzmann
constant, $u^{\alpha}$ the hydrodynamic three-velocity, $T$ the
temperature of the gas and $\mu_{e}$ the equilibrium chemical potential.
The particle number density is given by 
\begin{equation}
n=\int\left(\frac{u^{\alpha}p_{\alpha}}{c^{2}}\right)f^{(0)}dp^{*},\label{eq:n}
\end{equation}
where $dp^{*}=cd^{2}p/p_{0}$, and can be calculated as follows 
\begin{eqnarray}\nonumber
n=2\pi m^{2}c^{2}\frac{g_{s}}{h^{2}}\left[\int_{1}^{x_{F}}x\left(e^{-\left(\frac{\mu_{e}}{kT}-\zeta\right)}e^{\zeta\left(x-1\right)}+1\right)^{-1}dx \right. \\ 
\left. +\int_{x_{F}}^{\infty}x\left(e^{-\left(\frac{\mu_{e}}{kT}-\zeta\right)}e^{\zeta\left(x-1\right)}+1\right)^{-1}dx\right],\label{2dn}
\end{eqnarray}
where we have introduced the dimensionless variable $x=u^{\alpha}p_{\alpha}/mc^{2}$
as well the standard relativistic parameter $\zeta=mc^{2}/kT$ with
which the limiting cases can easily be identified.

Indeed, $\epsilon_{F}=mc^{2}(x_{F}-1)$ corresponds to the highest
energy level occupied at zero absolute temperature where all particles
fill the lower energy levels (up to such value) due to Pauli's exclusion
principle. It is worthwhile to point out that the number 1 that is
subtracted from $x_{F}$ corresponds to the rest energy. This allows
one to clearly isolate the relativistic limit which in the degenerate
case requires special care because of the interplay of the parameters
that characterize the system, particularly the energies \cite{HakimBook}.

\begin{figure}
\begin{center}
\includegraphics[width=0.4\textwidth]{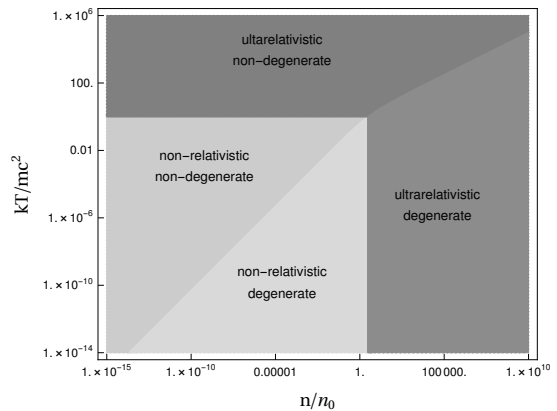} \caption{\label{fig:1} The phase diagram for a 2D fermions gas.}
\end{center}
\end{figure}

The degenerate relativistic behavior is characterized by $e^{\zeta-\mu_{e}/kT}\ll1$,
which implies that the distribution function on the first integral
of Eq. (\ref{2dn}) is bounded and therefore can be replaced by a
step-function, while the second integral becomes negligible. Therefore
the particle number density can be calculated in such scenario as
follows
\begin{equation}
n=2\pi m^{2}c^{2}\frac{g_{s}}{h^{2}}\int_{1}^{x_{F}}xdx=\pi m^{2}c^{2}\frac{g_{s}}{h^{2}}\left(x_{F}^{2}-1\right).\label{ndr}
\end{equation}
If we introduce the quantum 2D density $n_{0}=\pi m^{2}c^{2}g_{s}/h^{2}$,
then the relativistic Fermi energy can be written as a function of
the number density as
\begin{equation}
\frac{\epsilon_{F}}{mc^{2}}=\sqrt{\frac{n}{n_{0}}+1}-1.
\end{equation}
Notice that since the non-relativistic case is characterized by $e^{\zeta-\mu_{e}/kT}\gg1$,
by considering that for a degenerate gas $\mu_{e}\sim\epsilon_{F}$,
one can write the condition for a non-relativistic degenerate system
as $\epsilon_{F}\ll mc^{2}$. When this condition is introduced in
Eq. (\ref{ndr}), the non relativistic Fermi energy is recovered 
\begin{equation}
\frac{\epsilon_{F}}{mc^{2}}\simeq\frac{1}{2}\frac{n}{n_{0}}.
\end{equation}
All these cases are summarized in Fig. \ref{fig:1}, where the existence
of four regimes, corresponding to the inter-lapping of degenerate,
not degenerate, ultra relativistic and non-relativistic cases is shown.
This is in accordance with what is known for 3D electrons \cite{ETC}. 

The boundary of the non-relativistic and relativistic regimes is determined
by the relative magnitude of the rest energy to the thermal energy.
In particular, consider $mc^{2}$ compared with some factor of the
thermal energy, say $\alpha kT$. The constant $\alpha$ in the non-degenerate
case can be taken as $d/2$ by considering the value of the thermal
energy corresponding to the equipartition theorem. Thus, in the bidimensional
case here addressed, the line $T=mc^{2}/k$ separates non-relativistic
and relativistic regimes. However, as mentioned in \cite{SciRep2012},
relativistic effects can also be present at lower temperatures with
$\alpha=d+2=4$, and by considering also the effects of degeneracy
one can consider $\alpha=d+2+W\left((d+2)e^{-(d+2)}\right)\approx4.0684$,
where $W$ is the so-called Lambert-$W$ function. This yields a temperature
$T=mc^{2}/(\alpha k)$ separating non-relativistic and relativistic
regimes, together with $\epsilon_{F}=mc^{2}$, which in terms of density
is given by $n=3n_{0}$.

On the other hand, the boundary between the degenerate and non-degenerate
cases is given by the relative magnitude of thermal and Fermi energies,
i.e. $\alpha kT=\epsilon_{F}$, which can be expressed as follows
\begin{equation}
T=\frac{mc^{2}}{\alpha k}\sqrt{\frac{n}{n_{0}}+1}-1.
\end{equation}
In Fig. \ref{fig:1}, the particular case given by the equipartition
theorem was considered, since it gives the wider range in temperature
and density for the non-relativistic case. However, the effects of
considering the transition temperature have not yet been fully explored
\cite{SciRep2012}. It should be noted that the above expressions
can be written as a function of the dimension showing the influence
of the confinement on the characteristic energies of the system. 

\section{Balance equations and dissipative fluxes}

The kinetic equation described above (Eq. (\ref{eq:Uhlebeck})), allows
for the establishment of the hydrodynamic equations. Indeed, since
the state variables are given as moments of the distribution function,
the corresponding dynamics can be obtained through the evolution of
$f$, given by Eq. (\ref{eq:Uhlebeck}). In the relativistic case,
the state variables number density $n$ and internal energy $\varepsilon$
are related with the conserved fluxes through the following equations
\begin{equation}
n=\frac{1}{c^{2}}N^{\alpha}u_{\alpha}\label{eq:ene}
\end{equation}
\begin{equation}
n\varepsilon=\frac{1}{c^{2}}T^{\alpha\beta}u_{\alpha}u_{\beta}\label{epsilon}
\end{equation}
where $N^{\alpha}$ is the particle four-flux 
\begin{equation}
N^{\alpha}=\int fp^{\alpha}dp^{*},\label{eq:N-1}
\end{equation}
and $T^{\mu\nu}$ the energy-momentum tensor 
\begin{equation}
T^{\alpha\beta}=\int fp^{\alpha}p^{\beta}dp^{*}.\label{eq:T-1}
\end{equation}
As can be seen from the Eqs. (\ref{eq:ene}) and (\ref{epsilon}),
the description of the fluid is here given within Eckart's representation
(considering the fluid's frame) in order for the analysis and physical
interpretation to be more straightforward, particularly in the non-relativistic
limit \cite{KremerLibro,DeGroot,Eckart}. The hydrodynamic equations
are thus given by the conservation of both tensor quantities: 
\begin{equation}
N_{,\alpha}^{\alpha}=0,\label{eq:balance-1}
\end{equation}
\begin{equation}
T_{,\beta}^{\alpha\beta}=0\label{eq:balanceT-1}
\end{equation}
where here, and throughout this work, a comma indicates a covariant
derivative. When $N^{\mu}$ and $T^{\mu\nu}$ are calculated introducing
$f^{(0)}$ in Eqs. (\ref{eq:N-1}) and (\ref{eq:T-1}), one obtains
\begin{equation}
N_{(0)}^{\alpha}=2\pi m^{2}c^{2}g_{s}u^{\alpha}I_{1},\label{eq:Nf0}
\end{equation}
and 
\begin{equation}
T_{(0)}^{\alpha\beta}=2\pi m^{3}c^{4}g_{s}\left(I_{2}\frac{u^{\alpha}u^{\beta}}{c^{2}}
+\frac{1}{2}\left(I_{0}-I_{2}\right)h^{\alpha\beta}\right),\label{eq:Tf0}
\end{equation}
from which the dynamic equations in a local equilibrium, non-dissipative,
scenario are obtained, i. e. Euler's regime. Such equations can be
readily obtained and are here written as 
\begin{widetext}
\begin{equation}
\theta I_{1}+\frac{\mu_{e}u^{\beta}}{kT}\left[\frac{\mu_{e,\beta}}{\mu_{e}}\mathcal{I}_{1}-\frac{T_{,\beta}}{T}\left(\mathcal{I}_{1}+\frac{mc^{2}}{\mu_{e}}\mathcal{I}_{2}\right)\right]=0,\label{eq:n0-1}
\end{equation}
 \begin{equation}
h_{\alpha}^{\nu}u^{\beta}u_{,\beta}^{\alpha}-\frac{c^{2}\mu_{e}}{\left(3I_{2}-I_{0}\right)kT}h^{\nu\alpha}\left[\frac{\mu_{e,\alpha}}{\mu_{e}}\left(\mathcal{I}_{2}-\mathcal{I}_{0}\right)-\frac{T_{,\alpha}}{T}\left(\left(\mathcal{I}_{2}-\mathcal{I}_{0}\right)-\frac{mc^{2}}{\mu_{e}}\left(\mathcal{I}_{3}-\mathcal{I}_{1}\right)\right)\right]=0,\label{eq:u0-1}
\end{equation}
and 
\begin{equation}
\frac{1}{2}\theta\left(3I_{2}-I_{0}\right)+\frac{\mu_{e}u^{\beta}}{kT}\left[\frac{\mu_{e,\beta}}{\mu_{e}}\mathcal{I}_{2}-\frac{T_{,\beta}}{T}\left(\mathcal{I}_{2}-\frac{mc^{2}}{\mu_{e}}\mathcal{I}_{3}\right)\right]=0,\label{eq:e0-1}
\end{equation}
\end{widetext}
which correspond with the particle, momentum and energy balances,
respectively. Here $\theta=u_{,\alpha}^{\alpha}$ and the integrals
$I_{i}$ and $\mathcal{I}_{i}$ above are given by
\begin{equation}
I_{n}=\frac{1}{2\pi h^{2}}\frac{1}{m^{n+1}c^{2n+2}}\int_{-\infty}^{\infty}\frac{\left(u^{\alpha}p_{\alpha}\right)^{n}}{\Sigma+1}dp^{*},\label{eq:In}
\end{equation}
\begin{equation}
\mathcal{I}_{n}=\frac{1}{2\pi h^{2}}\frac{1}{m^{n+1}c^{2n+2}}\int_{-\infty}^{\infty}\frac{\left(u^{\alpha}p_{\alpha}\right)^{n}\Sigma}{\left[\Sigma+1\right]^{2}}dp^{*}.\label{eq:IIn}
\end{equation}
It is worthwhile to point out that Eqs. (\ref{eq:u0-1}) and (\ref{eq:e0-1})
are obtained by projecting Eq. (\ref{eq:balanceT-1}) in the direction
orthogonal and parallel to $u_{\alpha}$ respectively. More precisely,
(\ref{eq:u0-1}) is obtained by contracting Eq. (\ref{eq:balanceT-1})
with 
\begin{equation}
h^{\mu\gamma}=\eta^{\mu\gamma}-\frac{u^{\mu}u^{\gamma}}{c^{2}},\label{h}
\end{equation}
which is the so-called spatial projector and represents the direction
orthogonal to the time direction, given by $u^{\alpha}$. It is worthwhile
to recall the reader that in this representation space and time directions
are fixed by $h^{\alpha\beta}$ and $u^{\alpha}$ \emph{in the fluid's
comoving frame.}

On the other hand, by considering the Chapman-Enskog expansion, where
the distribution function is written as 
\begin{equation}
f=\sum_{n=0}^{\infty}f^{\left(n\right)},\label{CE}
\end{equation}
with $f^{\left(0\right)}$ being the local equilibrium solution described
in the previous section and $f^{\left(n\right)}$ the non-equilibrium
deviation to $n-th$ order in the Knudsen parameter \cite{KremerLibro,DeGroot,Ch-E},
the out of local equilibrium transport equations can be written to
any order in $n$. In particular, for $n=1$, the so-called Navier-Stokes
regime is obtained which features both thermal and viscous dissipation.
Indeed, in such a regime, thermal dissipation is characterized by
a heat flux which is given in terms of the energy-momentum tensor
as 
\begin{equation}
q^{\sigma}=h_{\alpha}^{\sigma}u_{\beta}T_{\left(1\right)}^{\alpha\beta},\label{qu}
\end{equation}
where 
\begin{equation}
T_{\left(1\right)}^{\alpha\beta}=\int f^{\left(1\right)}p^{\alpha}p^{\beta}dp^{*}.\label{te1}
\end{equation}
As mentioned in Section 1, the main goal of this work is precisely
to establish a general expression for the heat flux, given by Eq.
(\ref{qu}), and to calculate the corresponding transport coefficients
within a relaxation time approximation. In order to accomplish such
task, the following sections are devoted to the calculation of $T_{\left(1\right)}^{\alpha\beta}$
within Marle's model.

\section{First order non-equilibrium solution within Marle's model}

In order to establish dissipative fluxes, it is necessary to asses
the mathematical structure of the non-equilibrium solution to Eq.
(\ref{eq:Uhlebeck}), which is in general a highly involved task.
For this reason, model equations are usually introduced proposing
a simpler form of the collision kernel (r. h. s. of Eq. (\ref{eq:Uhlebeck})),
but retaining its most important basic characteristics. Moreover,
relaxation or BGK-like models \cite{BGK } turn the integrodifferential
equation into an algebraic one and allows one to readily obtain valuable
approximations for the transport coefficients that relate dissipative
fluxes with state variable's gradients.

In the relativistic case, the most widely employed relaxation models
are Marle's model \cite{Marle}, which has the issue of not being
able to treat massless particles, and the Anderson-Witting model \cite{AW1,AW2}
which is usually preferred in developments such as relativistic Lattice
Boltzmann. However, although such model represents a covariant formulation,
it corresponds to the so-called Landau-Lifshitz or energy frame where
addressing the relevant limits is somehow not so intuitive as in the
Eckart's or fluid frame here considered. Furthermore, the choice of
the appropriate frame for the description of a particular relativistic
processes can be a delicate task and it is thus crucial to consider
the previous observations, in particular when introducing a relaxation
parameter \cite{Takamoto,MendezAIP2010}.

In this section, we shall determine the first order in the gradients
non-equilibrium distribution function of a gas of electrons in 2D
by applying Marle's model 
\begin{equation}
p^{\alpha}\frac{\partial f}{\partial x^{\alpha}}=-\frac{m}{\tau}\left(f-f^{(0)}\right),\label{eq:marle}
\end{equation}
which is consistent with the fluid's frame representation here considered.
Although, as was said before, the Anderson-Witting model is the most
used to describe this kind of systems, this is not the first time
that a model of this type has been used to study relativistic degenerate
gases \cite{Kelly}. Moreover, there are yet ways to avoid its problems
and correct the model \cite{Takamoto,MendezAIP2010}. In Eq. (\ref{eq:marle})
$m$ is the particle's mass and $\tau$ is a parameter related with
the characteristic relaxation time for the distribution function.
In order to obtain the first order correction to the distribution
function due to the spatial gradients, we introduce $f=f^{\left(0\right)}+f^{\left(1\right)}$
as indicated above from which one obtains, for the Chapman-Enskog
method, the following expression 
\begin{equation}
f^{(1)}=-\frac{\tau p^{\alpha}}{m}\frac{\partial f^{(0)}}{\partial x^{\alpha}}.\label{eq:f1}
\end{equation}
where for the derivatives on the right hand side one makes use of
the local equilibrium assumption. Considering a representation where
the independent state variables are $\mu_{e}$, $T$ and $u^{\alpha}$
yields the following expression for $f^{(1)}$ 
\begin{equation}
f^{(1)}=\frac{\tau g_{s}}{mh^{2}}\frac{\Sigma~ p^{\mu}}{\left(\Sigma+1\right)^{2}}\left(\frac{T_{,\mu}}{T}\left(1-\frac{u_{\nu}p^{\nu}}{\mu_{e}}\right)-\frac{\mu_{e,\mu}}{\mu_{e}}+\frac{u_{\nu;\mu}p^{\nu}}{\mu_{e}}\right).\label{eq:f1-2}
\end{equation}
Introducing the distribution function up to first order Eq. (\ref{eq:f1-2})
into Eqs. (\ref{eq:N-1}) and (\ref{eq:T-1}) leads to 
\begin{eqnarray}\nonumber
N^{\alpha}=N_{(0)}^{\alpha}+\tau\frac{g_{s}}{mh^{2}}\frac{\mu_{e}}{kT}\int p^{\alpha}p^{\mu}\frac{\Sigma}{\left(\Sigma+1\right)^{2}}\times\\
\left(-\frac{\mu_{e,\mu}}{\mu_{e}}+\frac{T_{,\mu}}{T}\left(1-\frac{u_{\nu}p^{\nu}}{\mu_{e}}\right)+\frac{u_{\nu;\mu}p^{\nu}}{\mu_{e}}\right)dp^{*},\label{eq:Nf1}
\end{eqnarray}
and 
\begin{eqnarray}\nonumber
T^{\alpha\beta}=T_{(0)}^{\alpha\beta}+\tau\frac{g_{s}}{mh^{2}}\frac{\mu_{e}}{kT}\int p^{\alpha}p^{\beta}p^{\mu}\frac{\Sigma}{\left(\Sigma+1\right)^{2}}\times \\
\left(-\frac{\mu_{e,\mu}}{\mu_{e}}+\frac{T_{,\mu}}{T}\left(1-\frac{u_{\nu}p^{\nu}}{\mu_{e}}\right)+\frac{u_{\nu;\mu}p^{\nu}}{\mu_{e}}\right)dp^{*}.\label{eq:Tf1}
\end{eqnarray}
In order to obtain the constitutive equation for the heat flux, the
integrals in expression (\ref{eq:Tf1}) need to be evaluated. Such
procedure can be performed using the relations in Ref. \cite{HFBidi},
and leads to 
\begin{widetext}
\begin{eqnarray}
&T^{\alpha\beta}=T_{(0)}^{\alpha\beta}-\pi\tau g_{s}m^{3}c^{4}\zeta\left\{ \frac{\mu_{e}}{mc^{2}}\frac{\mu_{e,\mu}}{\mu_{e}}\left[\frac{2}{c^{2}}u^{\alpha}u^{\beta}u^{\mu}\mathcal{I}_{3}
-\left(h^{\alpha\mu}u^{\beta}+h^{\mu\beta}u^{\alpha}+h^{\alpha\beta}u^{\mu}\right)\left(\mathcal{I}_{3}-\mathcal{I}_{1}\right)\right]\right.\nonumber \\
&+\frac{T_{,\mu}}{T}\left[\frac{2}{c^{2}}u^{\alpha}u^{\beta}u^{\mu}\left(\mathcal{I}_{4}-\frac{\mu_{e}}{mc^{2}}\mathcal{I}_{3}\right)\right.-\left.\left(h^{\alpha\mu}u^{\beta}+h^{\mu\beta}u^{\alpha}+h^{\alpha\beta}u^{\mu}\right)
 \left(\left(\mathcal{I}_{4}-\mathcal{I}_{2}\right)-\frac{\mu_{e}}{mc^{2}}\left(\mathcal{I}_{3}-\mathcal{I}_{1}\right)\right)\right]\\ \label{eq:Tfinal}
&-\frac{1}{4}\left(h^{\alpha\mu}h^{\beta\nu}+h^{\alpha\beta}h^{\nu\mu}+h^{\mu\beta}h^{\nu\alpha}\right)u_{\nu,\mu}\left(\mathcal{I}_{4}-2\mathcal{I}_{2}+\mathcal{I}_{0}\right)
  \left.-\frac{1}{c^{2}}\left(u^{\alpha}u^{\gamma}u_{\nu,\gamma}h^{\beta\nu}+u^{\alpha}u^{\beta}u_{\nu;\mu}h^{\nu\mu}+u^{\gamma}u^{\beta}u_{\nu,\gamma}h^{\nu\alpha}\right)\left(\mathcal{I}_{2}-\mathcal{I}_{4}\right)\right\},\nonumber
\end{eqnarray}
\end{widetext}
where $\zeta=mc^{2}/kT$. Notice that here we wrote the momentum tensor
in its irreducible form in the 2+1 fluid frame representation: 
\begin{equation}
p^{\mu}=p_{\gamma}h^{\mu\gamma}+\frac{u^{\gamma}p_{\gamma}}{c^{2}}u^{\mu},\label{p}
\end{equation}
given by $u^{\alpha}$ as the temporal direction and $h^{\mu\gamma}$
(see Eq. (\ref{h})) representing the corresponding orthogonal plane.
Such decomposition allows for the separation of proper time and spatial
derivatives. The importance of such a step lies on the fact that the
existence of the Chapman-Enskog solution of order $n$ to the kinetic
equation relies on the assumption that the time derivatives of the
state variables, in principle given by $f^{(n)}$ itself, can be approximately
substituted by their expressions corresponding to the previous order
solution $f^{(n-1)}$.

In view of the arguments mentioned above, $u^{\mu}\left(\partial T/\partial x^{\mu}\right)$,
$u^{\mu}\left(\partial u^{\alpha}/\partial x^{\mu}\right)$ and $u^{\mu}\left(\partial\mu_{e}/\partial x^{\mu}\right)$
in Eq. (\ref{eq:Tfinal}) are now replaced by their expressions given
by the Euler equations (see Eqs. (\ref{eq:n0-1})-(\ref{eq:e0-1})),
that is:
\begin{widetext}
 \begin{equation}
\frac{u^{\alpha}}{T}\frac{\partial T}{\partial x^{\alpha}}=\theta\frac{kT}{mc^{2}}\frac{1}{\left(\mathcal{I}_{1}\mathcal{I}_{3}-\mathcal{I}_{2}^{2}\right)}\left(\mathcal{I}_{2}I_{1}-\frac{1}{2}\mathcal{I}_{1}\left(3I_{2}-I_{0}\right)\right),\label{tpunto}
\end{equation}
\begin{equation}
u^{\alpha}\frac{\partial\mu_{e}}{\partial x^{\alpha}}=-\frac{1}{\mathcal{I}_{1}}kT\theta\left[I_{1}+\left(\mathcal{I}_{2}-\frac{\mu_{e}}{mc^{2}}\mathcal{I}_{1}\right)\frac{1}{mc^{2}\left(\mathcal{I}_{1}\mathcal{I}_{3}-\mathcal{I}_{2}^{2}\right)}\left(\mathcal{I}_{2}I_{1}-\frac{1}{2}\mathcal{I}_{1}\left(3I_{2}-I_{0}\right)\right)\right],\label{mupunto}
\end{equation}
\begin{equation}
u^{\alpha}\frac{\partial u_{\mu}}{\partial x^{\alpha}}=-h_{\mu}^{\alpha}\frac{c^{2}}{kT\left(3I_{2}-I_{0}\right)}\left[\left(\mathcal{I}_{0}-\mathcal{I}_{2}\right)\frac{\partial\mu_{e}}{\partial x^{\alpha}}+\left[\left(mc^{2}\left(\mathcal{I}_{1}-\mathcal{I}_{3}\right)+\mu_{e}\left(\mathcal{I}_{2}-\mathcal{I}_{0}\right)\right)\right]\frac{1}{T}\frac{\partial T}{\partial x^{\alpha}}\right].\label{upunto}
\end{equation}
\end{widetext}
Once the substitution is carried out, $T^{\alpha\beta}$ is obtained
in terms of solely spatial gradients of the state variables and thus
the constitutive equation for the heat flux can be evaluated by means
of Eq. (\ref{qu}). Such a calculation is shown to some detail in
the next section.

\section{The heat flux}
As mentioned above, introduction of Eqs. (\ref{tpunto})-(\ref{upunto})
in Eq. (\ref{eq:Tfinal}) leads to an expression for the heat flux
by means of the projection
\begin{equation}
q^{\sigma}=h_{\alpha}^{\sigma}u_{\beta}T^{\alpha\beta},\label{eq:heatflux}
\end{equation}
As is well known, only forces and fluxes of same tensorial rank are
coupled in constitutive equations. Thus, the heat flux in this case
is only driven by $\mu_{e,\beta}$ and $T_{,\beta}$. Indeed, one
finds 
\begin{widetext}
\begin{eqnarray}
q^{\sigma}=-\pi\tau g_{s}m^{3}c^{6}\left\{ -h^{\sigma\beta}\frac{\mu_{e,\beta}}{\mu_{e}}\frac{\mu_{e}}{kT}\left(\left(\mathcal{I}_{3}-\mathcal{I}_{1}\right)+\frac{\zeta\left(\mathcal{I}_{2}-\mathcal{I}_{4}\right)\left(\mathcal{I}_{2}-\mathcal{I}_{0}\right)}{\left(3I_{2}-I_{0}\right)}\right)\right.\label{eq:heatWithForces}\\
+\left.h^{\sigma\beta}\frac{T_{,\beta}}{T}\left[\frac{\mu_{e}}{kT}\left(\left(\mathcal{I}_{3}-\mathcal{I}_{1}\right)+\frac{\zeta\left(\mathcal{I}_{2}-\mathcal{I}_{4}\right)\left(\mathcal{I}_{2}-\mathcal{I}_{0}\right)}{\left(3I_{2}-I_{0}\right)}\right)-\frac{\zeta^{2}\left(\mathcal{I}_{2}-\mathcal{I}_{4}\right)\left(\mathcal{I}_{3}-\mathcal{I}_{1}\right)}{\left(3I_{2}-I_{0}\right)}+\zeta\left(\mathcal{I}_{4}-\mathcal{I}_{2}\right)\right]\right\} ,\nonumber 
\end{eqnarray}
\end{widetext}
which can be written as

\begin{equation}
q^{\sigma}=-L_{T}h^{\sigma\beta}\frac{T_{,\beta}}{T}-L_{\mu}h^{\sigma\beta}\frac{\mu_{e,\beta}}{\mu_{e}},\label{eq:constitutive}
\end{equation}
where $L_{\mu}$ and $L_{T}$ are the transport coefficients associated
with heat conduction and can be written as 
\begin{eqnarray}\nonumber
L_{T}=&\tau g_{s}\pi m^{3}c^{6}\left\{ \frac{\mu_{e}}{kT}\left(\left(\mathcal{I}_{3}-\mathcal{I}_{1}\right)+\zeta\frac{\left(\mathcal{I}_{2}-\mathcal{I}_{0}\right)\left(\mathcal{I}_{2}
-\mathcal{I}_{4}\right)}{\left(3I_{2}-I_{0}\right)}\right)\right. \\
& \left. -\zeta\left(\mathcal{I}_{2}-\mathcal{I}_{4}\right)\left(\zeta\frac{\mathcal{I}_{3}-\mathcal{I}_{1}}{\left(3I_{2}-I_{0}\right)}-1\right)\right\} \label{lt}
\end{eqnarray}
\begin{eqnarray}
L_{\mu} =-\tau g_{s}\pi m^{3}c^{6} \frac{\mu_{e}}{kT}\left(\mathcal{I}_{3}-\mathcal{I}_{1}+\zeta\frac{\left(\mathcal{I}_{2}-\mathcal{I}_{0}\right)\left(\mathcal{I}_{2}-\mathcal{I}_{4}\right)}{\left(3I_{2}-I_{0}\right)}\right).\label{lmu}
\end{eqnarray}
Also notice that, using the following identity 
\[
I_{n}\left(n+1\right)=K+\zeta\mathcal{I}_{n+1}
\]
for $K$ a constant and $n=0,1,2,\dots$, the thermal conductivities are equal in magnitude, i. e. $L_{\mu}=-L_{T}$
and thus one can write a Fourier-like constitutive equation: 
\begin{equation}
q^{\sigma}=\kappa h^{\sigma\beta}\Theta_{,\beta}\label{eq:kappa}
\end{equation}
where $\kappa=L_{T}kT/\mu_{e}$ and $\Theta=\mu_{e}/kT$ can be thought
of as a generalized thermal force in a similar fashion as some authors
propose for the non-degenerate and non-relativistic case in 3D \cite{KremerLibro,Israel63}.
Figure 2 shows the thermal conductivity $\kappa$ in Eq. (\ref{eq:kappa})
as a function of $\zeta$ for different values of the chemical potential.
\begin{figure}
\begin{center}
\includegraphics[width=0.4\textwidth]{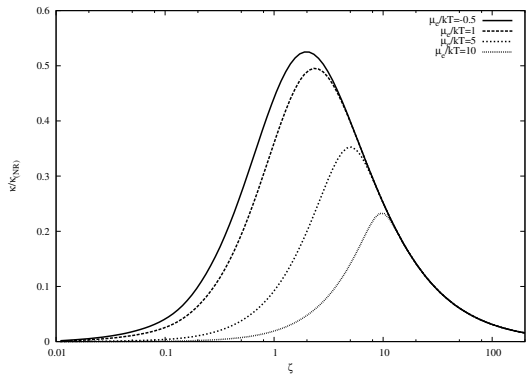} 
\end{center}
\caption{Thermal conductivity $\kappa/\kappa_{(NR)}$ as a function of $\zeta$
for different values of the chemical potential.}
\end{figure}

\section{Discussion of the results}

In order to compare the results obtained in the previous sections
with their 3D counterparts, we begin by rewriting the constitutive
equation for the heat flux, Eq. (\ref{eq:constitutive}), as follows
\begin{equation}
q^{\alpha}=\bar{\kappa}h^{\alpha\beta}\left(T_{,\beta}-\frac{T}{nh_{e}}p_{\beta}\right),\label{eq:kappaK}
\end{equation}
Thus, in order to qualitatively compare the present 2D results with
the values reported in Ref. \cite{KremerLibro}, the relevant conductivity
coefficient is given by
\begin{equation}
\bar{\kappa}=\frac{h_{e}\kappa}{kT^{2}},\label{eq:kappaGraphs}
\end{equation}
where $h_{e}$ is the enthalpy per particle:
\begin{equation}
 h_{e}=\varepsilon+\frac{p}{n}=\frac{mc^{2}}{2}\left(\frac{3I_{2}-I_{0}}{I_{1}}\right).
\end{equation}
Within this approximation the thermal conductivity depends linearly
in the relaxation parameter, which must be specified. However, the
dependence on temperature, or more precisely on the relativistic parameter
$\zeta$, has not been completely determined for bidimensional systems
\cite{UR2D,Coelho2,Coelho}.

As previously discussed in Refs. \cite{MendezAIP2010,Takamoto}, Marle's
model is proposed as a special relativistic generalization of its
non-relativistic counterpart, the BGK model. However, in order to
obtain the transport coefficients that coincide with the ones obtained
by solving the complete Boltzmann equation, in the relativistic scenario,
Marle's model must be modified. The transport coefficients obtained
by model equations are in general proportional to the parameter $\tau$
in Eq. (\ref{eq:marle}), that plays the role of a relaxation time
in the non-relativistic case, this time is proportional to the time
between collisions $\tau_{c}$, which for hard disks is given by 
\begin{equation}
 \tau_{c}=\frac{1}{bn\mathcal{V}},
\end{equation}
where $b$ is the diameter of the disks and $\mathcal{V}$ could be
the adiabatic sound speed $v_{s}$, the mean relative velocity $\left\langle g\right\rangle $
or the mean velocity $\left\langle v\right\rangle $. A modified parameter
$\tau$,which has been shown to yield more precise results is proposed
in Ref. \cite{MendezAIP2010}, is
\begin{equation}
\tau\sim\frac{1}{bn\mathcal{V}\left\langle x\right\rangle },\label{eq:tau-mod}
\end{equation}
where $\left\langle x\right\rangle $ is the mean value of the relativistic
parameter $x=u_{\alpha}p^{\alpha}/mc^{2}$. Here, the expression given
in Eq. (\ref{eq:tau-mod}) is considered, together with $\mathcal{V}=\sqrt{2}\left\langle v\right\rangle $
and the mean velocity being estimated as $\left\langle v\right\rangle =I_{v}/mc^{2}I_{0}$,
where 
\begin{equation}
\left\langle v\right\rangle =\frac{I_{0}^{-1}}{2\pi mc^{2}h^{2}}\int_{-\infty}^{\infty}\frac{\left|p\right|}{m}\left[\Sigma+1\right]^{-1}dp^{*},
\end{equation}
Also notice that $\left\langle x\right\rangle =I_{1}/I_{0}$.

\begin{center}
\begin{figure}
\begin{center}
\includegraphics[width=0.45\textwidth]{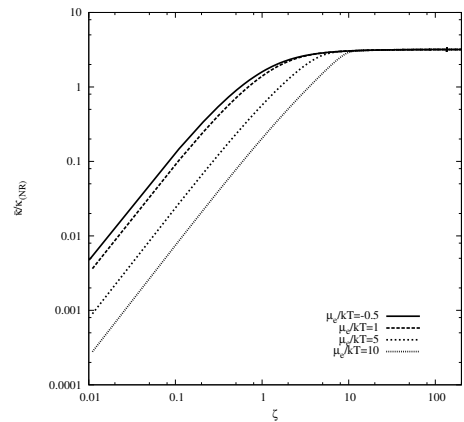} \caption{\label{fig:2} The dimensionless thermal conductivity $\bar{\kappa}/\kappa_{(NR)}$
as a function of $\zeta$ for different values $\mu_{e}/kT$.}
\end{center}
\end{figure}
\end{center}

\begin{figure}
\begin{center}
\includegraphics[width=0.45\textwidth]{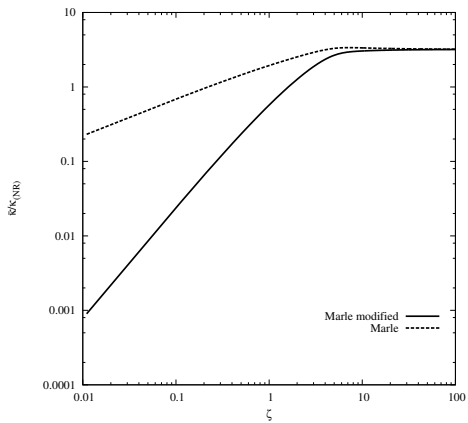} \caption{\label{fig:3} Comparison of the thermal conductivity with Marle and
modified Marle models for $\mu_{e}/kT=5$.} 
\end{center}
\end{figure}

The conductivity coefficient given in Eq. (\ref{eq:kappaGraphs})
for a gas of electrons is plotted in Fig. \ref{fig:2}, considering
the modified parameter $\tau$ as mentioned above. The normalized
(to $\sqrt{T}$) thermal coefficient is independent of the temperature
for large values of $\zeta$, consistent with the non-relativistic
and non-degenerate case. Notice also that the coefficient decreases
with $\mu_{e}/kT$ and that the result here obtained for a two dimensional
gas of electrons is qualitatively similar to the one reported in the
three-dimensional scenario in Ref. \cite{KremerLibro}. Meanwhile,
Fig. \ref{fig:3} shows the importance of considering the factor $\left\langle x\right\rangle $
introduced in $\tau$, which brings the curve of the coefficient to
resemble the results in the 3D case.
The non-relativistic and non-degenerate limit can be explored to detail
in an analytical fashion. Indeed, considering $\exp\left(-\mu_{e}+mc^{2}\right)/kT\gg1$,
one can approximate the integrals in (\ref{eq:In}) and (\ref{eq:IIn})
as follows 
\begin{equation}
I_{n}\approx\frac{1}{2\pi h^{2}}\frac{1}{m^{n+1}c^{2n+2}}\int_{-\infty}^{\infty}\frac{\left(u^{\alpha}p_{\alpha}\right)^{n}}{\Sigma}dp^{*},
\end{equation}
and also notice that in this limit $I_{n}=\mathcal{I}_{n}$. Thus
one has 
\begin{equation}
\bar{\kappa}\approx\frac{kc}{\sqrt{2}d}\left[-\frac{\exp\left(-\zeta\right)\left(3+6\zeta+2\zeta^{2}\right)}{\left(1+\zeta\right)^{3}K_{1}\left(\zeta\right)}\right]
\end{equation}
which, upon expansion of the Bessel function of the second kind $K_{1}\left(\zeta\right)$,
can be expressed as
\begin{equation}
 \kappa\approx2\frac{kc}{d\sqrt{\pi}}\zeta^{-1/2}\left[1-\frac{3}{8}\frac{1}{\zeta}+\dots\right]
\end{equation}
where the leading term is consistent with the classical non-relativistic
limit 
\[
\kappa_{NR}\approx\frac{2kc}{\sqrt{\pi}d}\frac{1}{\zeta^{1/2}}.
\]

\section{Final remarks}

In the present work, an analytical expression for a thermal conductivity
coefficient was established based on the relativistic kinetic theory
of gases within Marle's relaxation approximation and the Chapman-Enskog
expansion. It is worthwhile to comment on the fact that, in general,
the thermal coefficient to which one refers to as the thermal conductivity
depends on the thermodynamic forces that are considered as independent
sources of dissipation. In Refs. \cite{HFBidi,Garcia2019},
the heat flux is driven both by the temperature and density gradients
which are considered as independent forces in the representation where
$T$ and $n$ are taken as the scalar state variables. In the present
work however, since the gas dealt with is a quantum system, the chemical
potential replaces the number density as a relevant state variable.
In such a representation, both transport coefficients are identical
in magnitude which allows for the identification of a generalized
thermodynamic force, namely $\mu_{e}/kT$, and a single relevant transport
coefficient.

On the other hand the authors consider it valuable to emphasize the
importance of the role the particular value of the relaxation parameter
in BGK-like approximations plays in the magnitude and general behavior
of the transport coefficients, in particular for numerical simulations
as lattice Boltzmann methods. As is well known, these type of approximations
are extremely valuable in order to asses the general structure of
constitutive equations as they retain the force-flux relations as
obtained from the complete Boltzmann equation in most cases. However,
all the details of the collisions and thus the particular molecular
interaction, are included in one single parameter instead of various
collision integrals. The cost of turning the integrodifferential equation
in an algebraic relation is that of having to be extremely careful
when addressing the behavior of transport coefficients. In particular,
in the relativistic case two BGK-like models, namely Marle and Anderson-Witting,
have been proposed which in general lead to different approximations
for such quantities. Moreover, within those methods, the choice of
the characteristic parameter is not unique. In Marle's case, $\tau$
can be corrected from its expression in the non-relativistic case
in order to resemble more closely the value obtained both with the
complete kernel and Anderson-Witting model \cite{Takamoto,MendezAIP2010}.

The relevance and applications of the results here obtained can be
inferred from the comments included in the introduction and the references
cited therein. In this final section we would like to add that the
results here presented contribute to the understanding of the behavior
of bidimensional degenerate relativistic fluids. These systems, which
have been shown to have important and novel applications, have been
scarcely addressed in the literature from the theoretical point of
view. Moreover, the calculation of the viscosity coefficients in order
to complete the description of the dissipative properties of these
type of gases, is also lacking and will be the focus of a future publication.
As a final comment the authors wish to emphasize the importance of
addressing these type of calculations using the Chapman-Enskog method,
as here presented, and further considering the complete integral kernel,
which will be tackled in the near future. Indeed, as pointed out in
Ref. \cite{Coelho}, the Chapman-Enskog method leads to better results
than the simpler and more generally employed Grad moment method in
the case of relativistic hydrodynamics.

\end{document}